\documentclass[12pt]{article}

\usepackage{amsmath}
\usepackage{amssymb}

\usepackage{indentfirst}
\usepackage{amssymb}
\usepackage{amsfonts}
\usepackage{amscd}
\usepackage{amsbsy}
\usepackage{amsthm}
\usepackage{latexsym}
\usepackage{graphicx,color} 
\usepackage[dvipsnames]{xcolor}
\usepackage[colorlinks]{hyperref}
\hypersetup{linkcolor=blue,citecolor=blue,urlcolor=blue}
\def\nn{\nonumber}       
\def\beq{\begin{eqnarray}}
\def\eeq{\end{eqnarray}}
\def\ln{\,\mbox{ln}\,}

\DeclareMathOperator{\cx}{\square}

\def\al{\alpha}
\def\be{\beta}
\def\ch{\chi}

\def\de{\delta}

\def\la{\lambda}

\def\pa{\partial}

\def\rh{\rho}
\def\si{\sigma}

\def\ph{\varphi}

\def\th{\theta}

\def\Ga{\Gamma}
\def\De{\Delta}
\def\La{\Lambda}

\usepackage[symbol]{footmisc} 
\usepackage{float} 
\usepackage{cite}  

\begin{document}

\begin{center}
\renewcommand*{\thefootnote}{\fnsymbol{footnote}} 
{\Large
Bounce solutions with quantum vacuum effects of massive fields
\\
and subsequent Starobinsky inflation}
\vskip 6mm

{\bf Wagno Cesar e Silva}\,
$^{a}$
\hspace{-1mm}\footnote{E-mail address: \ wagnorion@gmail.com},
\,
{\bf Samuel W. P. Oliveira}\,
$^{b}$
\hspace{-1mm}\footnote{E-mail address: \ sw.oliveira55@gmail.com}
\;
and
\;
{\bf Ilya L. Shapiro}\,
$^{b, c}$
\hspace{-1mm}\footnote{E-mail address: \ ilyashapiro2003@ufjf.br}
\vskip 6mm

${a)}$ Centro Brasileiro de Pesquisas Físicas,
\\
Rua Dr. Xavier Sigaud 150, Urca, 22290-180, Rio de Janeiro, RJ, Brazil
\vskip 2mm

${b)}$ PPGCosmo, Universidade Federal do Espírito Santo,
\\
29075-910, Vit\'oria, ES, Brazil
\vskip 2mm
	
${c)}$ Departamento de F\'{\i}sica, ICE, Universidade Federal de Juiz de Fora,
\\
Campus Universitário, 36036-900, Juiz de Fora, MG, Brazil
\end{center}
\vskip 2mm

\begin{abstract}

\noindent
We extend the previous work about the cosmological solutions
with a bounce without modifications of gravity or introducing an extra
scalar field. The main finding was that the bounce is possible in
the initially contracting Universe filled with matter. After a strong
contraction, matter gains the equation of state close to the one of
radiation, such that the effect on matter on the evolution of the
FLRW metric disappears at the classical level. However, this effect
comes back owing to the quantum trace anomaly in the matter/radiation
sector. In the present contribution, we explore the weak impact of
massive fields on the anomaly-driven bounce solution and discuss
the role of the vacuum terms. The masses are assumed small and
regarded as small perturbations, which enables using trace anomaly
even in this case. On the other hand, by adding the $R^2$ term to
the action, we arrive at the model with the trans-Planckian bounce
and subsequent Starobinsky inflation. In such a framework,
using the numerical analysis, we consider three scenarios providing
bounce solutions. 

\vskip 2mm
	
	
\end{abstract}

\setcounter{footnote}{0} 
\renewcommand*{\thefootnote}{\arabic{footnote}} 

\section{Remembering Alexei Alexandrovich Starobinsky}
\label{secAlstar}

Before we present our contribution to this volume dedicated to 
the memory of Professor Starobinsky, let us tell a little bit about 
the history of contacts and collaboration with him by one of the 
authors of this paper — Ilya Shapiro, on whose behalf these 
reminiscences are conveyed below.

Alexei Alexandrovich was the person who greatly
impressed many people who had professional and personal
contacts with him and I was not an exception. We briefly met
with him before on several occasions, but the first time we had
serious conversation was at the meeting in Paris in 2006, devoted
to the 25 years of inflation. I am not a professional cosmologist and 
did not know well ``who is who'' in this community. The organizers
invited me because of our papers on the model of inflation based
on the conformal anomaly - something which was done originally
by Fischetti, Hartle and Hu \cite{fhh}. We added a few new details
and some people got interested, but not the main cosmologists 
present at this meeting.
At some point, Starobinsky addressed me and explained a few
points which we did not understand well. This proved very useful
for us since we got much better idea about the relation of anomaly
induced inflation and the Starobinsky model. In general, some 
discussions in this meeting were hot. What impressed me was that 
nobody ever complained if Starobinsky was saying something. 
It looked he was one of a very few participants who had this 
privilege.

Next time we met in 2008, when Starobinsky came to Juiz de Fora
for the first time for a small meeting which we organized as a
continuation of the larger conference in Jo\~ao Pessoa.
I have an impression that because of this visit Starobinsky started
to read about Minas Gerais and came to be interested in the
works of the artist of the XVIII - XIX centuries, called Aleijadinho.
This name belongs to an outstanding painter and sculptor, 
unfortunately not very much known out of Brazil.
Next time, Alexei Alexandrovich came for a few days and
asked to organize his visit to Ouro Preto or to Tiradentes, the
historic cities where one can see some Aleijadinho's works. In the 
next few years, we went to the region of Tiradentes, which is closer 
to Juiz de Fora, and everybody liked it a lot. Every time, in the next 
few visits, Starobinsky was contracting a particular driver to visit 
the town called Congonhas, where one can see the main sculptures 
of Aleijadinho: the set of Twelve Prophets sculptures made of a
soapstone between 1800 and 1805, and painted sculptures in the
famous Stations of the Cross scenes, carved between 1796 and 1799.
When we came to the same place with Alexei Alexandrovich later
on, on the way from Ouro Preto, he gave us detailed and very
interesting lecture about Aleijadinho and about these sculptures, 
which he knew at the
professional level and estimated very high. Another detail was that,
when we were walking in the Ouro Preto (old-style city, former
capital of Minas Gerais state), it looked like he knew every corner
and every direction, like as he would have a GPS in his head.
This and many other things of this sort impressed a lot all our
company, including me.

On top of these excursions, every time we had a lot of discussions
about Physics. We started with quantum effects in the early Universe
but, this stage, it was difficult to overcome some differences.
Anyway, at least for me, these discussions were very instructive.
At some point we decided to change the subject and I asked 
Starobinsky to look at the calculation of equivalence between $f(R)$ 
gravity and metric-scalar theory. I made this small calculation from 
scratch, without basic knowledge of the literature.
He looked at the calculations and immediately said that the same
method can be applied to many other theories. We decided to offer
this problem to one of my best students Felipe Salles. In the next
months, with Felipe we met some mathematical difficulties and
asked former Juiz de Fora student, Davi Rodrigues, to assist us. 
Eventually, after two years of work and a lot of discussions we 
could finish a paper with four authors \cite{fREddi}, based on this 
idea. Needless to say that
me and also our young collaborators were very excited to work
under the problem suggested by Starobinsky.

After this, our long discussions about quantum effects and higher
derivatives started to produce some results. Eventually, we did two
works. The first one was about the transition between stable and
unstable phases of the anomaly-induced inflation \cite{StabInstab}.
This was our main idea in the mentioned works of 2003, but we
did not know well how to connect it to cosmology because there 
was no reliable model of quantum corrections in the transition 
period. At some point, Alexei
Alexandrovich suggested to simplify things and explained to us
how the approximation works in this case. This lesson was very
useful for our group, and the approach appeared to be operational
in some other models. The second work emerged from numerous
discussions of how to deal with higher derivative models in
cosmology. The conventional approach (mainly in quantum theory) 
was the one by Simon \cite{Simon-90}. In our discussions, we 
always agreed that it is not a perfect thing, but at least enables us to 
ignore the problems. And at some point we decided to put it to 
practise, for the first time, that resulted in \cite{AnaR}.

Even after more than a year after Alexei Alexandrovich passed
away, it is difficult for me to accept this. Every time we have a
doubt about cosmology or many other issues, the first idea is to
ask the main expert in the field and it is painful to remember that
he is not available. We miss not only his knowledge, constructive
criticism and advises, but also his outstanding personality. What
is true is that Prof. Starobinsky left a strong imprint on many
people worldwide, including in Brazil and in our small group.
And many people, including us, will remember him again and again.
\vskip 3mm



\section{Introduction}
\label{sec0}

The Standard Cosmological Model \cite{Peeb93,Weinb08,BarTee12}
is a very successful theory but leaves significant theoretical and
observational challenges. Some of the issues may be addressed by
introducing the cosmic inflation phase in the early stage of the
evolution of the Universe \cite{star,Guth,Linde}. However, certain
conceptual limitations of this framework persist unresolved. In
particular, the existence of an initial singularity is unavoidable
in the framework of GR \cite{Penrose-sing,Hawking-sing}, and even
inflation may be insufficient to solve this problem. This fact may be
regarded as a strong motivation to modify GR, either through
introducing the scalar fields, modifying the gravitational action, or
introducing some exotic forms of  ``matter'' with an unusual equation
of state (see, e.g., \cite{Coles}).

In this framework, bouncing models emerge as an interesting
alternative to avoid singularities and, in some cases, provide an
explanation for the primordial density perturbations. There are many
different models of cosmological bounce, where the contraction of
the Universe precedes its expansion. The systematic study of these
models attracted a lot of interest in recent decades  (see, e.g.,
the reviews  \cite{Novello2008,Peter}). In
particular, the possibility of achieving nonsingular bouncing
solutions within the framework of the anomaly-induced
effective action was explored in \cite{AnoBo21} by two of
the present authors (earlier, the bounce in such a model was
discussed in \cite{anju}). In a subsequent work \cite{BoRa24},
an analytic solution describing the cosmological bounce
was found without additional \textit{ad hoc} assumptions,
i.e., without modifying the action of gravity or introducing a
scalar field, or accounting for vacuum quantum effects.

The scheme of bounce owing to the quantum trace anomaly in the
radiation sector \cite{AnoBo21} is quite simple, as it is based on
the one-loop approximation\footnote{
It is worth mentioning that
matter-bouncing cosmological models, including with the discussion
of primordial black holes and induced gravitational waves production,
were elaborated in the \cite{Papanikolaou2024} (partially based on the
techniques of \cite{Banerjee2022}). However, these
models are not based on the quantum corrections.}.
The last means, the violation of local
conformal symmetry is restricted to the terms linear in the
scale parameter $\si = \log\, a(t)$. The model does not require
\textit{ad hoc} modifications of gravity and ignores vacuum quantum
effects which may be relevant, in principle. On the other hand, this
model of bounce leaves several open questions.
The list starts with the numerical discrepancy between ignoring the
vacuum (purely gravitational, i.e. metric-dependent) part of anomaly
and the fact that the bounce requires the trans-Planckian energy
density at the point of minimum of $a(t)$ with vanishing $H=0$.
The way out consists of looking beyond the basic QED framework
and invoking the possible non-perturbative effects, i.e., including
the higher orders in $\si(t)$.  The second problem concerns the
analysis of cosmic perturbations which may be critical for the
consistency of the bounce model \cite{Peter}. Finally, there are
several smaller problems which we shall address in the present
contribution. The first one is to check how the small masses of the
quantum fields affect the bounce solution and its stability in both
vacuum- and radiation-dependent anomaly-induced cases. In what
follows, we deal with this problem in a way similar to what was
done in the modified Starobinsky model of inflation
\cite{asta,StabInstab}. Another question is what happens to the
bounce solution if we add the $R^2$ term with a large coefficient
to the classical action of gravity, as it is effectively done in the
Starobinsky model \cite{star}. If the bounce is not critically
affected by such an addition, one can assume that the vacuum
contributions to the anomaly are irrelevant compared to the
dominating $R^2$-term. In this way, we arrive at the model with
cosmological bounce at very high energies and Starobinsky
inflation occurs later on, after a significant expansion of the
Universe.

The paper is organized as follows.  Sect.~\ref{sec1} is a brief
survey of the previous work \cite{BoRa24} where the cosmological
bounce solution based on the trace anomaly in the radiation sector
of the theory has been found. In Sect.\,\ref{sec2}, we review
the approach of \cite{Shocom,asta}, which permits to use the
anomaly-induced effective action in the case of massive fields.
This approach is extended to the presence of radiation and is
used a basis to explore the dynamics of the scale factor in the FLRW
(Friedmann-Lemaitre-Robertson-Walker) cosmological metric.
Sect.~\ref{sec3} describes the stability analysis of the bounce
solution under linear perturbations of the conformal factor, taking
into account the effects of massive fields. In Sect.~\ref{sec4}, we
present the numerical results for the semiclassical bounce and
Sect.~\ref{secA} describes the extended version of the
radiation-dominated bounce with an additional $R^2$-term, which
may have a subsequent phase of the Starobinsky inflation. Finally,
in the last section, we draw our conclusions.

\section{Bounce solution from the trace anomaly}
\label{sec1}

Let us start with a brief survey of the previous work.
The model considered in \cite{BoRa24} consists solely of
the radiation sector from the anomaly-induced action of the
metric $g_{\mu\nu} = \bar{g}_{\mu\nu} \exp\,\si(t)$, with
the flat fiducial metric $\bar{g}_{\mu\nu} = \eta_{\mu\nu}$
at $a(t) = a_0$.
The gravitational action may include the cosmological term.
We use the signature $(+,-,-,-)$ for the Minkowski metric and
adopt the natural system of units, where $c= 1$ and $\hbar = 1$.
The action with the anomaly-induced term is \cite{rie,RadiAna}
\beq
\label{action}
\Ga \,=\,
-\,\dfrac{1}{16\pi G}\int d^4x\sqrt{-g}\,\big(R+2\La\big)
\,-\,
\dfrac{\,\be g^{2}}{4\,}\int d^4x\sqrt{-\bar{g}}\,\bar{F}^2\si .
\eeq
The bars denote quantities defined using the fiducial
metric, e.g., $\bar{F}^2 =
\bar{g}^{\mu\al}\bar{g}^{\nu\be}F_{\mu\nu}F_{\al\be}$,
and $\be g^4$ is the one-loop beta function for the square of
the gauge coupling $g^{2}$,
\beq
\be \,=\,-\,
\frac{2}{(4\pi)^2}\,\Big(
\frac{11}{3}\,C_1 - \frac{1}{6}\,N_{cs}
- \frac{4}{3}\,N_f\Big).
\label{ym19}
\eeq
Here $N_{cs}$ and $N_f$ are the numbers of complex scalars
and fermions coupled to the given vector field, and $C_1$ is
the Casimir operator of the corresponding gauge group.

The anomalous term in (\ref{action}) produces the modified
Friedmann equation,
\beq
H^2 \,=\, \frac{a_0^4}{a^4}\,\bigg[
\frac{\rho_{r0}}{M_{P}^{2}}
\,+\, \frac{2 \pi \be g^2{\bar F}^2}{3M_P^2}\,
\log \Big(\frac{a}{a_0}\Big)\bigg]
\,+\, \frac{\La}{3}\,,
\label{Hubble}
\eeq
where $M_P=1/\sqrt{G}$ is the Planck mass and $\rho_{r0}$ is a
radiation energy density at $a(t) = a_0$. The expression (\ref{Hubble})
admits vanishing Hubble parameter for a negative $\be$ and enables
one to estimate the energy density of the radiation at the bounce
point, when the scale factor assumes its minimum value. Thus, the
bounce occurs owing to the equilibrium between the classical
radiation term and the quantum correction in the
radiation-gravitational sector. In the analysis of \cite{BoRa24}, the
cosmological constant is supposed to satisfy $0 <  \La \ll M_P^2$, in
this case its presence does not change the qualitative result stated
above.

\section{Anomaly-induced action of massive fields}
\label{sec2}

Since the typical energy scale in the region of the bounce is,
presumably, very high, all matter fields can safely be regarded
massless. Then, relevant quantum effects near a bounce point
may be described by the effects of massless fields in the
zero-order approximation. On the other hand, it is interesting
to check whether the masses of quantum fields may affect the
bounce solution described in the previous section (see \cite{AnoBo21}
for more details), including the stability in the contraction phase.
The bounce occurs in the far UV when the masses of real fields
filling the Universe are completely irrelevant because the equation
of state is one of radiation. Thus, we need to evaluate the
effect of masses in the quantum corrections to the terms in the
action (\ref{action}). Since the masses are relatively small, the
most useful technique is the conformal approach developed in
\cite{Shocom,asta}.

Let us start with a brief discussion of deriving the conformal
anomaly and the anomaly-induced effective action in the theory
with small masses of the fields. It is worth noting that there are
many possible definitions of an anomaly for the classically
non-conformal theories (see, e.g., \cite{Sebastian-2024} and
references therein). Our approach is focused on effective
action and treats the masses of the quantum fields as small
perturbations \cite{PoImpo}.

In the high-energy regime of our interest, we can assume that
only scalar and fermion fields are massive. Let us denote these
masses by $m_{s}$ and $m_{f}$ respectively. To get rid of these
dimensional parameters, one can use the approach that can
be called conformal St\"{u}ckelberg trick. The masses of the
fields, the Newton constant, and the cosmological constant are
replaced by the powers of a new auxiliary scalar field $\ch$, i.e.,
\beq
m_{s,f} \rightarrow \frac{m_{s,f}}{M}\ch,
\qquad
\frac{1}{16\pi G}R \rightarrow \frac{M_P^{2}}{16\pi M^{2}}
\big[R\ch^{2}
+6(\pa\ch)^{2}\big],
\qquad
\La \rightarrow \frac{\La}{M^{2}}\ch^{2},
\label{mass}
\eeq
where $M$ is a new dimensional parameter. Applying this
procedure to the Einstein-Hilbert term provides the local conformal
invariance in all sectors of the theory. The new field $\chi$ is
auxiliary and should not be quantized. It is assumed that at some
point we replace back $\chi \rightarrow M$, but this will be done
only after we derive the effective action.

To guarantee the local conformal invariance, we have to
assume that $\ch$ transforms as
\beq
\ch \,=\, \bar{\ch}\,e^{-\si(x)},
\eeq
while the metric and matter fields transform as usual, i.e.,
\beq\label{conf_transf}
g_{\mu\nu} \,=\, \bar{g}_{\mu\nu}\,e^{2\si(x)},
\qquad
\ph \,=\, \bar{\ph}\,e^{-\si(x)},
\qquad
\psi \,=\, \bar{\psi}\,e^{-3\si(x)/2},
\qquad
A_\mu \,=\, \bar{A}_\mu.
\eeq


The relevant part of the classical action that is a subject of
renormalization is formed by the vacuum sector, including the
Einstein-Hilbert and the cosmological constant terms,
four-derivative terms, plus the matter sector, with a radiation,
scalar and fermionic terms,
\beq
S \,=\, \int d^4x\sqrt{-g}\,\bigg\{
-\dfrac{1}{16\pi G}\big(R+2\La\big) +
 a_1 C^2 + a_2 E_4 + a_3 \cx R \bigg\}
+S_{\textrm{matter}}\,.
\label{higher}
\eeq
Here $a_i$ are dimensionless coefficients, $C^2
= R^{2}_{\al\be\mu\nu} - 2 R^{2}_{\al\be} + \dfrac13 R^2$
is the square of the Weyl tensor and
$E_4 = R_{\mu\nu\al\be}^2 - 4 R_{\al\be}^2 + R^2$
is the integrand of the topological Gauss-Bonnet term.
Under the replacement trick (\ref{mass}) the first two terms
become conformal invariant and the same for the matter action.

At the quantum level, the local conformal symmetry is violated
by the corresponding anomaly. Then, the effective action of the
background fields can
be derived using the approach of the anomaly-induced action
(see, e.g., \cite{OUP} for a detailed introduction). As we have
already mentioned, $\ch$ has to be treated as a background field,
similar to the metric. How does the trace anomaly change under
the procedure (\ref{mass})? The gauge invariance prevents
mass-dependent contributions to the radiation terms thus the
last term in the expression \eqref{action} does not get modified.
Thus, the conformal anomaly takes the form elaborated in the
previous works \cite{Shocom,asta},
\beq
\langle \mathcal{T}\,\rangle
\,=\,
-\bigg\{ wC^2+\frac{1}{4}\be g^{2}F^2+bE_4+c\,\square R
+\frac{M_P^2}{4\pi M^2}\tilde{f}\Big[R\ch^2+6(\pa\ch)^2\Big]
+\frac{M_P^4}{4\pi M^4}\tilde{g}\ch^4
\bigg\}.
\label{anomaly}
\eeq
The coefficients $w$, $b$ and $c$ are the semiclassical one-loop beta
functions for the higher-derivative terms in the vacuum sector. These
beta functions are well-known to depend on the number of fields of
different spins (see, e.g., \cite{birdav} or the textbook \cite{OUP}),
\beq
&&
w \,\,=\,\, \frac{1}{120(4\pi)^2}\,\big(
N_s + 6N_{f} + 12 N_v \big),
\nn
\\
&&
b \,\,=\,\, -\,\frac{1}{360(4\pi)^2}\,\big(
N_s + 11N_{f} + 62N_v \big),
\nn
\\
&&
c \,\,=\,\, \frac{1}{180(4\pi)^2}\,\big(
N_s + 6N_{f} - 18N_v \big).
\label{wbc}
\eeq
On the other hand, the beta functions for the Einstein-Hilbert and
cosmological terms are proportional, respectively, to the
dimensionless quantities \cite{Shocom}
\beq
\tilde{f}
\,=\,\frac{1}{3\pi M_P^{2}}\sum_{\textrm{fermions}}
N_{f}\,m_{f}^{2}\,,
\qquad
\tilde{g}
\,=\,\frac{1}{4\pi M_P^{2}\La}\Big(
\sum_{\textrm{scalars}}N_{s}\,m_{s}^{4}
-4\sum_{\textrm{fermions}}N_{f}\,m_{f}^{4}\Big),
\label{fg}
\eeq
where $N_{s}$ and $N_{f}$ are multiplicities of scalars
and fermions with the corresponding masses.

Since the $F^2$-part and the terms that depend on $\chi$
have the same conformal properties as the square of the Weyl
tensor, such contributions can be incorporated into a
conformally invariant structure $Y(g_{\mu\nu},A_{\mu},\chi)$,
and the integration procedure of the anomaly is reduced to the
usual case  \cite{rie,frts84}. In this way, after obtaining the
anomaly-induced effective action $\Ga_{\textrm{ind}}$
and fixing the conformal unitary gauge $\bar{\ch}\,e^{-\si}=M$,
one can write complete effective action for massive fields,
including the vacuum sector, in the form
\beq
&&
\Ga
\, = \,
S_{vac}+\Ga_{\textrm{ind}}[\ch \rightarrow M]
\,=\,S_{HD}
\,+\, S_c[{\bar g}_{\mu\nu},M]
 \,-\, \frac{3c+2 b}{36}\,\int d^4 x\sqrt{-g}\,R^2
\nn
\\
&&
\qquad \quad
+\,\,\int d^4 x \sqrt{-{\bar g}}\,\bigg\{
\si\Big(w\bar{C}^2-\dfrac{1}{4}\be g^{2}{\bar F}^2\Big)
+b\si \Big(\bar{E}-\frac23 \bar{\Box}
{\bar R}\Big)+2b\si{\bar \De}_4\si
\nn
\\
&&
\qquad  \quad
-\,\,\frac{e^{2\si}}{16\pi G}\Big[\bar{R}
+6(\pa\bar{\si})^2\Big](1-\tilde{f}\si)
-\dfrac{\La\,e^{4\si}}{8\pi G}(1-\tilde{g}\si)\bigg\},
\label{Eq_full}
\eeq
where $ \Delta_4=e^{-4\si}\bar{\Delta}_4$ is the Paneitz
operator \cite{FrTs-superconf,Paneitz},
\beq
\Delta_4\,=\,
\cx^2
+ 2R^{\mu\nu}\nabla_{\mu}\nabla_{\nu}
- \dfrac{2}{3}R\square
+\dfrac{1}{3}(\nabla^{\mu}R)\nabla_{\mu}.
\label{Paneitz_op}
\eeq
In Eq.\,\eqref{Eq_full}, the functional $S_c[{\bar g}_{\mu\nu},M]$
is an ``integration constant'' that, in the massless case and with the
homogeneous and isotropic background metric, does not contribute
to the dynamics of the conformal factor. In principle, the same
result cannot be guaranteed since this term is not conformally
invariant. This means, \eqref{Eq_full} should be treated as an
approximate expression. What is the physical sense of this
approximation? It is easy to note that (\ref{Eq_full}) is a local
version of the renormalization-group improved classical action
of the theory, where the global parameter of metric rescaling is
replaced by the local conformal factor. Thus, the approximations
we made are essentially the same as in the case of the renormalization
group based on the Minimal Subtraction scheme. The comparison
with explicit expressions shows that this approximation is reliable
in the UV \cite{apco,OUP}, when the masses of quantum fields
can be regarded as small corrections
compared to the energy scale of the curvature tensor or the
density of matter. In the case of a bounce based on the action
\eqref{action} this is exactly what we need.

Our purpose is to evaluate the effect of the mass terms related
to the coefficients $\tilde{f}$ and $\tilde{g}$ in Eqs.~(\ref{fg}), to
the two scenarios that provide nonsingular bouncing solutions:
\
$(i)$ when the complete anomaly-induced corrections are taken
into account, in particular the higher-derivative sector. This is
the case studied in \cite{AnoBo21}, for massless fields;
\
$(ii)$ the model \eqref{action} described by the mixing of
radiation and gravity in the effective action. In this case, the
higher-derivative terms are neglected, and the bounce is due
to the contribution of the radiation sector.
\
$(iii)$ The same as the previous, but with an extra $R^2$ term.

\subsection{Dynamic equations and de Sitter-like solutions}
\label{sec2.1}

Consider the spacetime described by the FLRW metric,
\beq
ds^{2}=N^{2}(t)dt^{2}-a^{2}(t)
\Big(
\frac{dr^{2}}{1-kr^{2}}
+r^{2}d\th^{2}
+r^{2}\sin^{2}\th\,d\ph^{2}
\Big),
\label{BImetric}
\eeq
where $a(t)=e^{\si(t)}$ is the scale factor of the universe,
$N(t)$ is the lapse function and $k$ describes the spatial
curvature. Since our interest is to study the dynamics of the
scale factor, $N(t)$ is employed as an auxiliary function to
obtain the second Friedmann equation.

By introducing the conformal time $d\eta = dt/a(t)$,
Eq.~\eqref{BImetric} admits the conformal transformation
\eqref{conf_transf} with the fiducial metric
\beq
\bar{g}_{\mu\nu} \,=\,
\textrm{diag}\Big(N^2(\eta),\,-\,\frac{1}{1-kr^{2}},-\,r^{2},
-\,r^{2}\sin^{2}\th\Big).
\eeq
Using this metric in the complete action \eqref{Eq_full}, we
derive the modified first Friedmann equation by applying
the variational principle with respect to the scale factor $a(\eta)$.
This procedure is equivalent to taking the trace of the modified
Einstein equations with the terms from the higher derivative
and radiation sectors. The results for the traces can be presented
in terms of $\si(t)=\ln a(t)$, taking $N=1$ and changing to the
physical time $t$,
\beq
T_{EH}\,+\,T_{HD}\,=\,T_{rad}\,,
\label{T-eq}
\eeq
where the traces of the equations of motion for the
corresponding terms in the action are
\beq
&&
T_{EH} \,=\,
-\,\dfrac{3M_P^2}{4\pi}\bigg[
\big(\ddot{\si}+2\dot{\si}^2+ke^{-2\si}\big)
(1-\tilde{f}\si)-\dfrac{\tilde{f}}{2}
\big(\dot{\si}^2+ke^{-2\si}\big)
-\dfrac{2\La}{3}\bigg(
1-\tilde{g}\si -\dfrac{\tilde{g}}{4} \bigg)\bigg],
\nn
\\
&&
T_{HD} \,=\,
6c\,\bigg[\ddddot{\si}
+7\dddot{\si}\dot{\si}
+4\ddot{\si}^2
+4\bigg(3-\frac{b}{c}\bigg)\ddot{\si}\dot{\si}^{2}
-\frac{4b}{c}\dot{\si}^4
-2k \bigg(1+\frac{2b}{c}\bigg)
\big(\dot{\si}^2+\ddot{\si}\big)e^{-2\si}\bigg],
\nn
\\
&&
T_{rad} \,=\,
-\,\dfrac{\be F^2}{4}e^{-4\si},
\label{EMM}
\eeq
with $k=0$ or $k=\pm 1$ for different space geometries.
We derive the
$00$-component, $\rh_i$, using the conservation law for
$T_{EH}$, $T_{HD}$ and $T_{rad}$,\footnote{The same
equation can be obtained by varying the action \eqref{Eq_full}
with respect to the lapse function $N(t)$ and assuming $N=1$
after that. See more details in \cite{Stud,RadiAna} and
\cite{Tensors}.}
\beq
\frac{d\rho_i}{da^3}
\,+\,\frac{4}{3}\frac{\rho_i}{a^3}
\,=\,\frac{T_i}{3a^3},
\label{density equation}
\eeq
whose general solution is given by
\beq
\rho_i(a) = C(a)\,a^{-4},
\quad
\mbox{with}
\quad
\frac{d C}{dt} = T_i\,a^3\,\dot{a}.
\label{coef. equation}
\eeq
After integrating the expression \eqref{coef. equation} for the
components \eqref{EMM} and collecting the results from each
sector, we arrive at the following equation
\beq\label{EoM}
\rho_{EH}
+\rho_{HD}
= \rho_{rad},
\eeq
where
\beq
&&
\rho_{EH}
\,=\,
\dfrac{3M_P^2}{8\pi}\bigg[\bigg(\dot{\si}^2
+k\,e^{-2\si}\bigg)(1-\tilde{f}\si)
-\dfrac{\La}{3}(1-\tilde{g}\si)\bigg],
\nn
\\
&&
\rho_{HD}
\,=\,
6b\dot{\si}^4
+6k\,(2b+c)\dot{\si}^2\,e^{-2\si}
+3c\,\big(\ddot{\si}^2-2\,\dddot{\si}\,\dot{\si}
-6\,\ddot{\si}\,\dot{\si}^2\big),
\nn
\\
&&
\rho_{rad}
\,=\,
\dfrac{1}{4}(\rho_{r0}+\be\bar{F}^2\si)\,e^{-4\si}.
\label{EOS}
\eeq

In the case of massless matter fields, i.e., $\tilde{f}=\tilde{g}=0$,
with $k=0$, and omitting the radiation term, Eq.\,\eqref{EoM}
reduces to a biquadratic equation for the Hubble parameter,
$H=\dot{\si}=const$, with the following dS-like solutions
(obtained in \cite{asta} from the trace term)
\beq
H_{\textrm{dS}}\,=\, \pm\,\frac{M_P}{\sqrt{-32\pi b}} \,
\bigg(\,1\pm \,
\sqrt{1+\frac{64\pi b}{3}\,\frac{\Lambda }{M_P^2}\,}
\bigg)^{1/2}\,.
\label{H}
\eeq
The solutions with the positive $H$ correspond to expanding,
and the ones with the negative $H$, to contracting Universes.
Let us note that, for $\La \ll M_P^2$, the $(+,-)$ version of
\eqref{H} gives the classical de Sitter solution with $H_1$, while
the  $(+,+)$  solution $H_2$ is close to the exponential one obtained
in \cite{mamo,star,FrTs84}. This solution corresponds to the
equilibrium between the quantum anomaly-induced part and the
classical part, i.e.,
\beq
H_1\,=\,\sqrt{\frac{\Lambda }{3}}
\qquad
{\rm and}
\qquad
H_2\,=\, \frac{M_P}{\sqrt{- 16\pi b}}\,.
\label{HH}
\eeq

On the other hand, even in the case of massive matter fields,
we can still find an approximate solution for $\si(t)=\ln a(t)$.
This is achieved by replacing \cite{asta},
\beq
M_P^2 \,\,\rightarrow \,\,\tilde{M}_P^2 \,=\,
M_P^2\big(1-\tilde{f}\si\big),
\qquad
\La \,\,\rightarrow\,\, \tilde{\La} \,=\,
\La\big(1-\tilde{g}\si\big),
\label{aprox.solut.}
\eeq
in the solutions for the massless case. From the second
solution in \eqref{HH}, one can easily integrate
$\tilde{H} = H_0\,(1-\tilde{f}\si)^{\frac{1}{2}}$,
under the initial condition $\si(0)=0$, to obtain, in the absence
of a cosmological constant and assuming $\tilde{g}=0$,
\beq
\si(t) \,=\, H_2\,t \,-\, \dfrac{H_2^2}{4}\tilde{f}t^2.
\label{sigma.solut}
\eeq
The parabola \eqref{sigma.solut} reproduces the numerical solution
\cite{Shocom} with the precision of $10^{-6}$. Unlike the basic
conformal case, this approximate solution for the conformal factor
leads to a running Hubble parameter that decreases linearly with time
and smoothly transforms into the usual \cite{star} Starobinsky
inflationary solution \cite{StabInstab}.

\section{Stability analysis}
\label{sec3}

As the first contribution, in this section, we consider the bounce
solution in the theory where the bounce is produced by vacuum
anomaly-induced terms. We extend the previous
consideration of the stability \cite{asta} to the contraction phase
of the bounce solution and also include, for the sake of generality,
the case of a theory with nonzero
cosmological constant and its logarithmic (i.e., proportional to
$\si(t)$-parameter) running according to the effective action
(\ref{Eq_full}). Indeed, according to the well-known results
\cite{BludRud} we can assume that the cosmological constant
is small, even in the very early Universe. Another relevant detail
is that the physically justified way to perturb the cosmological
solution is to consider the variation of $\si(t)$ and \textit{not}
the variation of $a(t)$. 
This issue was discussed in detail in
 \cite{asta} and the interested reader may look at this reference.
This difference is also described in the mathematical
literature \cite{cezare}. Anyway, it is worthwhile to explain 
it in more detail.
  
Let us consider the asymptotic stability of the solution 
$a(t) = \exp\big[\sigma (t)\big]$, perturbing either $a(t)$ 
or $\sigma (t)$. For the sake of simplicity, consider 
exponential time dependence, i.e.,
\beq 
a(t) \,=\,e^{H_0t} + \de a(t)
\qquad 
\mbox{and} 
\qquad 
\sigma(t) \,=\, H_0t + \de \sigma(t)\,,
\label{asigma}
\eeq
which obviously means 
\beq
\de a(t) \,=\, \,e^{H_0t}\de \sigma(t)\,.
\label{asigma-dif}
\eeq
This formula explains the difference. If the eigenvalues of 
the characteristic equation for the   $\sigma (t)$ are all 
negative, then the solution is stable in this variable. On the 
other hand, if at least one of these eigenvalues is larger 
than $-H_0$,   the same solution is unstable in the variable
$a(t)$.  It is important that, for the 
non-exponential form of $a(t)$, the different may still take 
place, but only for finite intervals of time, not asymptotically.
Still, this may be relevant because the mentioned interval may 
be greater than the duration of the corresponding epoch of 
the evolution of the Universe.

To explore the asymptotic stability conditions in the framework
described above, let us perform a linear perturbation of the
conformal factor, $\si(t)\rightarrow\si(t)+x(t)$, in the
Friedmann equation \eqref{EoM}, neglecting the radiation part
and assuming $k=0$. This leads to the following equation for
$x(t)$,
\beq
&&
\dot{\si}\dddot{x}
-\big(\ddot{\si}-3\dot{\si}^2\big)\ddot{x}
+\bigg(\dddot{\si}
+6\ddot{\si}\dot{\si}
-\frac{4b}{c}\dot{\si}^3\bigg)\dot{x}
\nn
\\
&&
\qquad
\quad
-\,\dfrac{M_P^2}{16\pi c}\,\bigg[
2\dot{\si}\big(1-\tilde{f}\si\big)\,\dot{x}
-\bigg(\dot{\si}^2\tilde{f}-\frac{\tilde{g}}{3}\La\bigg)x\bigg]
= 0.
\label{analytical.no.rad.}
\eeq
In the first approximation, one can consider that the perturbations
have frequencies much greater than the Hubble parameter
characterizing the dynamics of the background solution. This means,
we can consider $\tilde{H}=\dot{\si}=constant$ on a sufficiently
small interval of time $(t_i,t_i+\De t_i)$. We note that even using
this approximation, evaluating the stability analytically is a
challenging task when mass contributions are taken into
account. For the simplest case with $\La=\tilde{g}=0$, this implies
that $(1-\tilde{f}\si) = 1 $ and $\tilde{H}=H_0$. In this case, the
equation for the perturbation $x(t)$ reduces to a simpler form,
\beq
\dddot{x}
+3H_0\ddot{x}
-\dfrac{2b}{c}H_0^2\,\dot{x}
-\dfrac{b}{c}\,H_0^3\tilde{f}\,x
= 0,
\label{pert.eq.}
\eeq
whose solution can be written as
\beq
x(t) = C_0\,e^{\al t}\cos(\be t+\ph)
+ C_1\,e^{\la_3 t},
\label{kind.solut.}
\eeq
where $\la_{1/2}=\al\pm i\be$ and $\la_3$ are the roots of
the polynomial equation corresponding to \eqref{pert.eq.}.
In the expansion phase, the asymptotic stability corresponds
to a negative real part of $\la_{1/2}$, i.e., $\al<0$. The explicit
form of $\al$, $\be$, and $\la_3$ is
\beq
&&
\al
\, = \,
-H_0\bigg[1-\dfrac{1}{3\cdot2^{2/3}}
\Big(\dfrac{Z_1}{Z_3}
-\dfrac{Z_3}{2^{2/3}}\Big)\bigg],
\nn
\\
&&
\be
\, = \,
H_0\bigg[\dfrac{1}{\sqrt{3}\cdot2^{2/3}}
\Big(\dfrac{Z_1}{Z_3}
+\dfrac{Z_3}{2^{2/3}}\Big)\bigg],
\nn
\\
&&
\la_3
\, = \,
-H_0\bigg[1+\dfrac{2^{1/3}}{3}
\Big(\dfrac{Z_1}{Z_3}
- \dfrac{Z_3}{2^{2/3}} \Big)\bigg],
\eeq
where we used notations
\beq
&&
Z_1 =
-3\,\Big(3+\dfrac{2b}{c}\Big),
\qquad
Z_2 =
-\,54\,\Big[1+\dfrac{b}{c}
\Big(1-\dfrac{\tilde{f}}{2} \Big)\Big],
\nn
\\
&&
\quad
\;\textrm{and}\;
\quad
Z_3 =
\Big[Z_2+\sqrt{4\,Z_1^3+Z_2^2}\Big]^{1/3}.
\nn
\eeq
As in the inflationary case \cite{Shocom,asta}, it is possible to
classify the bounce models with respect to the particle contents
which affects the coefficients (\ref{wbc}). The classification is
based on the sign of coefficient\footnote{We
assume the value that emerges in all
regularizations except dimensional and covariant Pauli-Villars
\cite{anomaly-2004}, also see the recent discussion of this
issue in conformal quantum gravity \cite{WeylBoxR}.} $c$.
E.g., for the particle content of the Minimum
Supersymmetric Standard Model (MSSM), with
$N_{s,f,v} = (104, 32, 12)$, and the parameter
$\tilde{f}=10^{-3}$ \cite{asta}, we can solve
numerically the polynomial equation for
\eqref{pert.eq.} and obtain the results:
\beq\label{stab_c}
\la_{1/2} \approx -\big[1.50 \pm i\,(3.57)\big]H_0
\qquad
\textrm{and}
\qquad
\la_3 \approx -5.00 \times 10^{-4}H_0.
\eeq
In this case, stability is verified for the expansion phase,
i.e., for $H_0 > 0$, as discussed in previous works
\cite{asta}. On the other hand, the same formulas
demonstrate the instabilities for the contraction phase when
$H_0 < 0$. These values for $\lambda_{1/2}$
coincide with the analytical result for the massless case
obtained in \cite{AnoBo21}, where the contraction phase
was also analyzed. This happens because the effect of
$\tilde{f}$ is only significant for $\lambda_3$. From the
numerical side, to get a small magnitude of the roots
$\lambda_{1/2}$, it is necessary to have $\tilde{f} \geq 0.1$.
This contradicts the limit on this parameter imposed above and
also appears to be a non-realistic condition because it means that,
according to Eqs.~(\ref{fg}),
the spectrum of the underlying semiclassical GUT-like model
includes particles with the almost-Planck magnitude of masses.

For the particle content corresponding to the Minimal Standard
Model (MSM), where $N_{s,f,v} = (4, 24, 12)$, the oscillatory
part becomes negligible, and the set of eigenvalues consists
of positive and negative real roots. This feature signals an unstable
solution for both the expansion and contraction phases.

To complete our numerical analysis, consider the non-zero
values of $\Lambda$ and $\tilde{g}$.
Starting from the expression \eqref{analytical.no.rad.}, and
assuming the approximation of $\tilde{H}$ as constant,
we can obtain
\beq
&&
\tilde{H}\dddot{x}
+3\tilde{H}^2\ddot{x}
-\frac{4b}{c}\tilde{H}^3\dot{x}
-\dfrac{M_P^2}{16\pi c}\,\bigg[
2\tilde{H}\dot{x}
-\bigg(\tilde{H}^2\tilde{f}-\frac{\tilde{g}}{3}\La\bigg)x\bigg]
= 0,
\label{analytical.no.rad.2}
\eeq
where we use that $\big(1-\tilde{f}\si\big)=1$ and
$\big(1-\tilde{g}\si\big)=1$. Here, $\tilde{H}$ assumes the
standard structure of \eqref{H}. Finally, considering the MSSM
particle content and the estimates of $\tilde{g}=10^{-2}$ and
$\Lambda=10^{-12}$ \cite{Ana09}, we arrive at
\beq
\la_{1/2} \approx 1.46 \pm i\,(3.47)
\qquad
\textrm{and}
\qquad
\la_3 \approx 4.85 \times 10^{-4},
\qquad
\textrm{for}
\;\;
\tilde{H}<0.
\eeq
This result shows that for nonzero $\tilde{g}$ and $\Lambda$, the
solution corresponding to the contraction phase remains unstable
under small perturbations. This means the nonzero cosmological
constant and its running do not change the stability properties of
the bounce solution.

\section{Numerical results for the bounce solutions}
\label{sec4}

In this section, we present the numerical results for the scale
factor and the Hubble parameter in the two bounce scenarios,
$(i)$ and $(ii)$, mentioned in Section \ref{sec2}.
To this end, we use the expression for the trace equation
\eqref{T-eq} together with the constraint equation
\eqref{density equation}, which provides the energy balance
and can be useful for setting the initial conditions, especially
in the second scenario $(ii)$, based on the radiation effects.
The analysis is numerical and all the plots shown
in this section were generated using Mathematica \cite{Wolfram}.

\begin{figure}[ht!]
\centering
\includegraphics[scale=0.615]{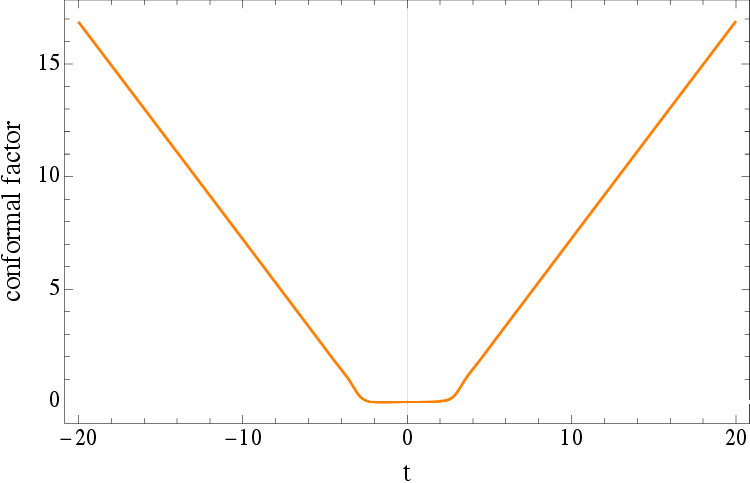}
\,
\includegraphics[scale=0.634]{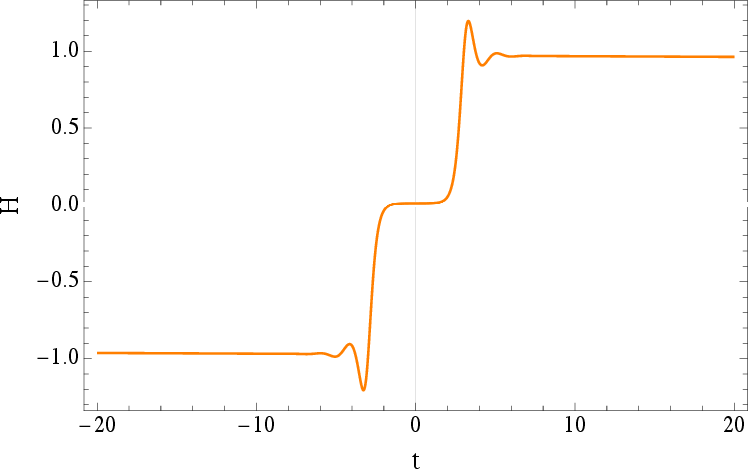}
\vspace{-5mm}
\begin{quotation}
\caption{\small	
Numerical solutions for the conformal factor $\si(t)$ in
scenario $(i)$. Here, we assumed the initial conditions:
$\si(0)=0, \ \dot{\si}(0)=-10^{-2}\tilde{H}, \ \ddot{\si}(0)=0,
\ \dddot{\si}(0)=0$. The right plot shows the Hubble
parameter $H(t)$.}
\label{fig1}
\end{quotation}
\end{figure}

For the first scenario, without loss of generality, we consider
$T_{EH} + T_{HD} = 0$ with $k = 0$. The solutions obtained in
this case are presented in Fig.~\ref{fig1}, where we adopt Planck
units for time $t$. Similar to the massless case studied in
\cite{AnoBo21}, it is also possible to find bouncing solutions in
the present context for specific initial conditions. Moreover, such
solutions persist under small variations of the initial conditions.

Fig.~\ref{fig1} contains only the configuration with
$\tilde{f}=10^{-3}$, $\tilde{g}=10^{-2}$ and $\La=10^{-8}$.
We tried some other configurations, especially
$a)$ $\tilde{f}=0$, $\tilde{g}=10^{-2}$ and $\La=10^{-8}$
\ and \
$b)$ $\tilde{f}=10^{-3}$, $\tilde{g}=0$ and $\La=10^{-8}$. All these
three configurations provide very similar behavior for the conformal
factor and the same is true for other sets of parameters. Indeed, the
mentioned values of $\tilde{f}$, $\La$, and $\tilde{g}$ are huge
compared to the physically acceptable ones. For instance, the
well-motivated bound for the first parameter is $\tilde{f} < 10^{-6}$,
which means the masses of the quantum particles do not exceed
$M_X \sim 10^{16} GeV$. The values we tried are far above this
bound and still do not produce significant changes. The same is
true for the parameters $\La$ and $\tilde{g}$. Thus, we conclude
that, in general, the effects of the masses of quantum fields do
not bring relevant changes to the bounce point in the scenario $(i)$.
Compared to the massless case, there are only small differences in
the shapes of the plots, but no qualitative changes. To be precise,
minimally visible changes are observed only in the presence of the
nonzero cosmological constant  $\La$ and at times far from the
bounce point, $t < -50$ and $t > 50$. From the physical side, the
negligible effect of masses should be expected since the closer one
gets to the transition point, the higher the typical energy becomes,
and the less relevant the mass effects of the fields are.
\begin{figure}[ht!]
\centering
\includegraphics[scale=0.611]{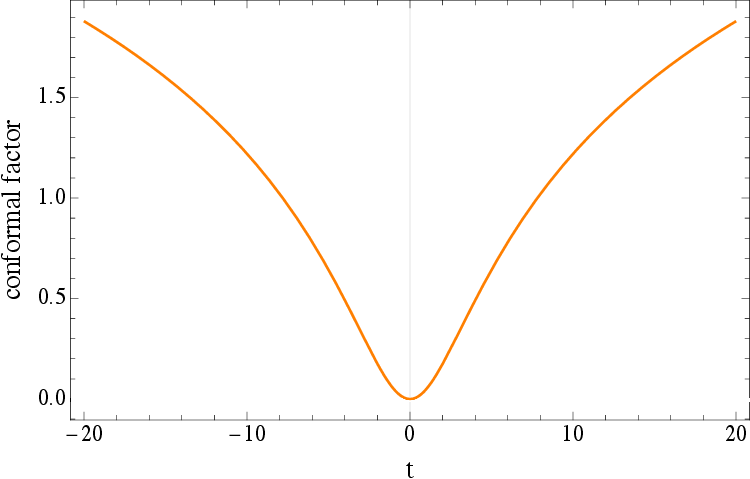}
\,
\includegraphics[scale=0.635]{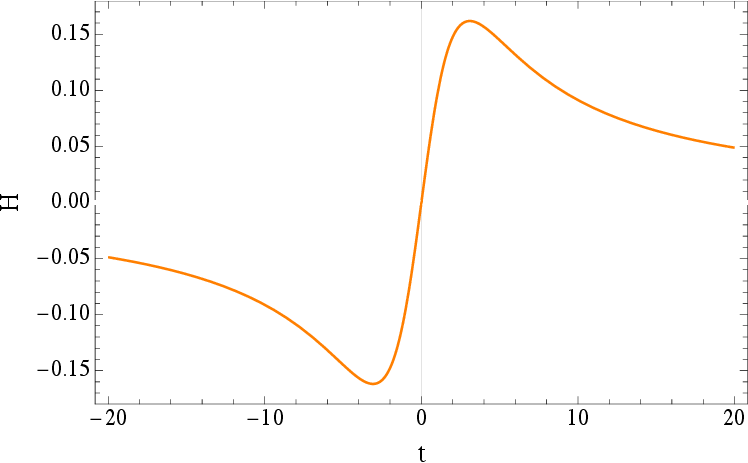}
\vspace{-5mm}
\begin{quotation}
\caption{\small
Numerical solutions for the conformal factor $\si(t)$ in scenario
$(ii)$, where the higher-derivative terms are neglected. We
assumed the value $\be g^{2}\bar{F}^2=0.1$ in the Planck units and
the initial conditions $\si(0)=0$ and $\dot{\si}(0)=-10^{-1}\tilde{H}$.
The right plot shows the Hubble parameter $H(t)$.}
\label{fig2}
\end{quotation}
\end{figure}

In the scenario $(ii)$, the equation for the trace is given by
$T_{EH} - T_{rad} = 0$, where we also assume $k = 0$. Fig.~\ref{fig2}
shows the corresponding solutions, with
$\tilde{f}=10^{-3}$, $\tilde{g}=10^{-2}$ and $\La=10^{-8}$.
In this case, we find that the quantum effects of massive fields
are even more subtle than in the first scenario.
The solutions with $\Lambda = 0$ and $\tilde{g} = 0$ do not exhibit
any visible difference between the massive and massless cases,
such that there is no reason to include the corresponding plots
in our presentation. To get a visible (still very small) difference with
respect to the massless case, it is necessary to postulate a large
magnitude of the cosmological constant, which contradicts the
standard assumptions \cite{BludRud}.

\section{The radiation-dominated bounce with the $R^2$-term}
\label{secA}

Let us, finally, present a third scenario providing a bounce
solutions. The action we shall consider here is composed
of the Einstein-Hilbert term, including the cosmological constant,
the $R^2$-term, and the radiation part induced by the anomaly,
\beq\label{actionR2}
\Ga_{R^2}
\,=\,
\int d^4x\sqrt{-g}\,\bigg[
-\dfrac{1}{16\pi G}\big(R+2\La\big)
+a_4R^2\bigg]
-\dfrac{\be g^{2}}{4}\int d^4x\sqrt{-\bar{g}}\,\bar{F}^2\si,
\eeq
where $a_4\approx 5\times 10^{8}$ is the huge coefficient of the
$R^2$-term required in the Starobinsky model \cite{star,star83}.
Taking a direct variational derivative of the action or, equivalently,
from the trace of the modified Friedmann equations, we arrive
at the following expression describing the dynamics of the
conformal factor,
\beq
\label{eomR2}
\ddddot\si
+7\dddot\si\dot\si
+4\,\ddot\si^2
+12\,\ddot\si\dot\si^2
+\dfrac{M_P^2}{48\pi a_4}\Big(\dot\si^2
+\frac{\ddot\si}{2}-\frac{\La}{3}\Big)
-\dfrac{\be g^{2}}{288\,a_4}\bar{F}^2e^{-4\si}
\,=\,0.
\eeq
Without the radiation term, we can easily reduce the order of
Eq.~\eqref{eomR2} and arrive at the same results as in \cite{star},
close to what we described in the first sections. In particular,
assuming
$\dot{\sigma}=H=constant$, there is the classical de Sitter
solution $H_1=\sqrt{\Lambda/3}$. In comparison with the
scenario $(i)$, the higher-derivative sector of \eqref{actionR2},
composed only of $R^2$, does not generate a term of the
type $\dot{\sigma}^4$ in the equation for the trace, and this
makes it more difficult to obtain a nontrivial analytical solution
in the de Sitter case.

\begin{figure}[h!]
\centering
\includegraphics[scale=0.9]{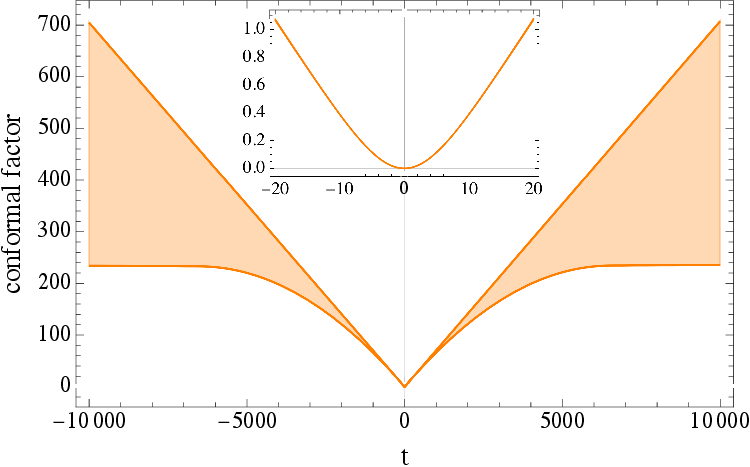}
\vspace{-2mm}
\begin{quotation}
\caption{\small	
Set of numerical solutions for the conformal factor $\sigma(t)$
in scenario $(iii)$, which includes the contribution from the
$R^2$-term.
The extreme curves are obtained by assuming the initial 
conditions: $\si(0)=0, \ \dot{\si}(0)=10^{-4}, 
\ \ddot{\si}(0)=10^{-2}, \ \dddot{\si}(0)=0$.
The smaller plot shows a zoom in the region around the
bounce point, for the interval $-20 \leq t \leq 20$ in
Planck units.
}
\label{fig3}
\end{quotation}
\end{figure}

On the other hand, we can explore the solution space of the scenario
$(iii)$ numerically by varying the parameters and initial conditions
and, in this way, find the bounce-type solutions, as presented in
Fig.\,\ref{fig3}.
The lower extreme curve represents the situation with a very small
coefficient $a_4$, approaching the well-known solution from
\cite{BoRa24}, where $a_4 = 0$.
While the upper extreme curve corresponds to the case with
$a_4 \approx 5 \times 10^{8}$. The region between these two
extreme solutions contains a set of possible bounce solutions with
different values for the parameters $a_4$, $\be g^{2} \bar{F}^{2}$,
and $\La$, as well as variations in the initial conditions.
Additionally, the smaller plot provides a zoom into the region
around the bounce point, i.e., in the higher energy region.
Note that the difference in the bounce solutions appears only
at the lower energies, but still above the Planck scale.
In this case, we observe that such solutions arise even without
the radiation term provided that fine-tuning of the initial
conditions are implemented.

\begin{figure}[h!]
\centering
\includegraphics[scale=0.87]{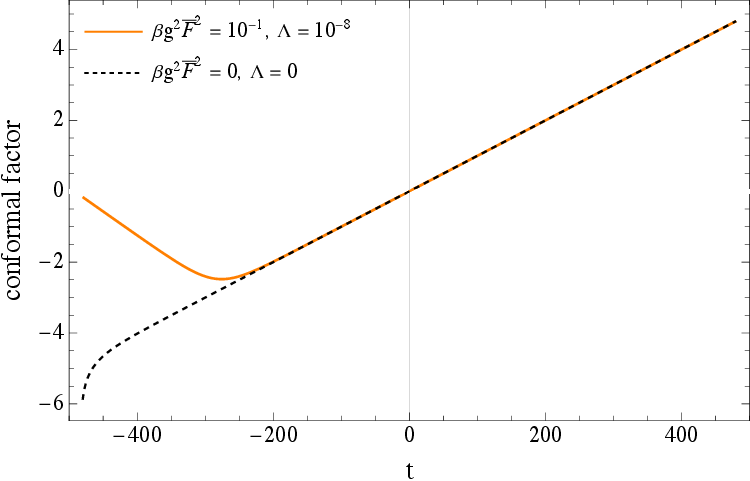}
\vspace{-2mm}
\begin{quotation}
\caption{\small	
Numerical solutions for the conformal factor $\si(t)$ in
scenario involving a large coefficient for the $R^2$-term
($a_4\approx 5\times 10^{8}$).
We assumed the initial conditions: $\si(0)=0,
\ \dot{\si}(0)=10^{-2}, \ \ddot{\si}(0)=0,\ \dddot{\si}(0)=0$,
and the interval $-480 \leq t \leq 480$ in the Planck units.
}
\label{fig4}
\end{quotation}
\end{figure}

However, it is important to mention that the bounce solutions
that do not take into account the radiation contribution are more
sensitive to changes in the initial conditions, indicating possible
dynamic instability.
For example, consider the numerical results for the conformal
factor, using the two configurations for $\be g^2 \bar{F}^2$
and $\La$, shown in Fig.\,\ref{fig4}.
Here, it is easy to see that the bounce is possible only in the
presence of the radiation term. This solution holds for both
zero and nonzero cosmological constant cases.
For a systematic study of dynamic instability in this system,
it is necessary to implement cosmological perturbations,
and this is an issue that we intend to address in future work.

\section{Conclusions and discussions}
\label{sec5}

We explored several particular problems related to the bounce
solutions in the gravity theory with quantum corrections. Assuming
that the Universe is in the initial contracting phase, it has to come
to the instant when the typical distance becomes very small and, therefore,
the typical energy scale is very high. At this point, the masses of
the particle contents of the Universe become irrelevant and the
effective equation of state is the one of radiation. On the other
hand, the quantum effects of effectively massless fields may be
described by the trace anomaly. The anomaly-induced effective
action may generate a cosmological bounce either by taking into
account the vacuum (purely metric-dependent) effects (model $i$)
or changing the equation of state of the radiation (model $ii$).

In the first place, we considered the effect of small masses of the
quantum fields on the anomaly-induced bounce. In both cases of
the models $(i)$ and $(ii)$, the masses do not produce strong
changes on the bounce solutions. The second issue is the stability
of the model $(i)$ that was explored for the first time. It turns out
that all versions of this model manifest instabilities in the
contracting phase. This output means that model $(i)$ cannot be
seriously considered as a prospective model to describe cosmological
bounce, at least in the form this model was formulated in
\cite{anju,AnoBo21}. Finally, we considered adding the $R^2$-term
to the most elegant and simplest bounce model $(ii)$.
The main problem of this model is that in the perturbative regime
\cite{BoRa24}, the bounce occurs in the deep trans-Planckian regime,
where we are not allowed to ignore vacuum quantum (and even the
quantum gravity) effects. The possible way out is to add the $R^2$
classical term with a numerically big coefficient, such that the
vacuum effects are shadowed by this dominating term. We have
found that, in this case, there is a bounce solution. However, the
stability properties in the presence of the dominating $R^2$ term
are difficult to establish because the investigation of this issue
requires an extensive numerical analysis for a variety of initial
conditions. We hope to complete this study in a future work.

\section*{Acknowledgments}

The authors are very grateful to Patrick Peter for discussions
and to Nicolas Bertini for the important assistance with the
preparation of plots presented in Sec.~\ref{secA}.
W.C.S. is grateful to \textit{Conselho
	Nacional de Desenvolvimento Cient\'{i}fico e Tecnol\'{o}gico}
(CNPq - Brazil) for supporting his postdoctoral project under
the PCI grant 302494/2024-3.
S.W.P.O expresses his gratitude to \textit{Funda\c{c}\~{a}o
	Coordena\c{c}\~{a}o de Aperfei\c{c}oamento de Pessoal de N\'{\i}vel
	Superior} (CAPES-Brazil) for supporting his PhD project.
The work of I.Sh. is partially supported by CNPq under the grant
305122/2023-1.



\end{document}